\documentclass[twocolumn,showpacs,preprintnumbers]{revtex4}
\usepackage{amssymb}

\usepackage{amsmath}
\usepackage{graphicx}
\usepackage{dcolumn}
\usepackage{bm}

\input{tcilatex}

\begin{document}

\title{A General Phase Matching Condition for Quantum Searching Algorithm}
\author{$^{1}$Che-Ming Li}
\author{$^{2}$Chi-Chuan Hwang}
\author{$^{3}$Jin-Yuan Hsieh}
\author{$^{1}$Kuo-Shong Wang}
\affiliation{$^{1}$Department of Mechanical Engineering, National Central University,
Chung-li, Taiwan.}
\affiliation{$^{2}$Department of Engineering Science, National Cheng Kung University,
Tainan, Taiwan.}
\affiliation{$^{3}$Department of Mechanical Engineering, Ming Hsin Institute of
Technology,Hsinchu, Taiwan.}

\begin{abstract}
A general consideration on the phase rotations in quantum searching
algorithm is taken in this work. As four phase rotations on the initial
state, the marked states, and the states orthogonal to them are taken
account, we deduce a phase matching condition for a successful search. The
optimal options for these phase are obtained consequently.
\end{abstract}

\pacs{ 03.67. Lx, 03.65.Bz}
\maketitle

Quantum mechanical algorithms have recently become more and more popular in
the field of computation science because they can speed up a computation
when compared with classical algorithms. A famous example is the factorizing
algorithm discovered by Shor\cite{shor}. Another, which is what we intend to
deal with in this work, is the quantum search algorithm developed by Grover %
\cite{grover1}\cite{grover2}. If there is an unsorted database containing $N$
items, and out of which only one marked item satisfies a given condition,
then using Grover's algorithm one will find the object in $O(\sqrt{N})$
quantum mechanical steps instead of $O(N)$ classical steps. In many
following researches it has been pointed out that the Grover algorithm is
optimal. Benett et al.\cite{optimal1} showed that no quantum algorithm can
search an object better than $O(\sqrt{N})$ steps. Boyer et al.\cite{optimal2}
gave tight bounds on Grover's algorithms. Zalka \cite{optimal3} has improved
the tight bounds to show that Grover's algorithm is optimal and proposed a
further improvement on the algorithm. Basically, Grover's algorithm consists
of three successive unitary operations. They are (1) the Walsh-Hadamard
transformation on the qubit $\left| 0\right\rangle $ to create an initial
state in the superposition of the basis states; (2) the $\pi $-angle
rotation of the marked state; and (3) the inversion of the initial state.

Although Grover\cite{grover3} also proposed that the Walsh-Hadamard
transformation can be replaced by almost any unitary transformation to
create the initial state, in this work, however, we will focus on the phase
rotations for the initial, the marked states, and the states orthogonal to
them. Long et al. [8] have shown that when arbitrary rotations for the
marked and the initial states are used instead of the inversions in Grover's
original algorithm, the rotation phases must satisfy certain matching
conditions. Galindo and Mart\'{\i}n-Delgado \cite{galindo} recently have
addressed a more general viewpoint on the phase rotations. They gave four
parameters in the Grover kernel since the rotations are operated on the four
states including the initial, the marked, and the two states orthogonal to
the former two. As the $\pi $-radian inversions for the initial and the
marked states were considered, Galindo and Martin-Delgado concluded that the
other two rotation phases must be equal. In this work, we will also study
the four phase rotations in the Grover kernel but derive a general matching
condition between the parameters without fixing any of them in advance.

Suppose we have to search $M$ objects out of $N$ unsorted elements, or in
expressions:

\begin{eqnarray}
f(w_{i}) &=&1\text{ , }i=1,2,...,M. \\
f(w_{i}) &=&0\text{ , }i=M+1,M+2,...,N\text{.}  \notag
\end{eqnarray}%
To begin the search, first construct a space spanned by the orthonormal set $%
\left| w_{i}\right\rangle $, $i=1,2,...,N$, and give the evenly distributed
state as our initial state:

\begin{eqnarray}
\left| s\right\rangle &=&\frac{1}{\sqrt{N}}\sum_{i=1}^{N}\left|
w_{i}\right\rangle \\
&=&\sqrt{\frac{M}{N}}\left| \underset{\symbol{126}}{w}\right\rangle +\sqrt{%
\frac{N-M}{N}}\left| \underset{\symbol{126}}{r}\right\rangle \text{,}  \notag
\end{eqnarray}%
where

\begin{eqnarray*}
\left| \underset{\symbol{126}}{w}\right\rangle &=&\frac{1}{\sqrt{M}}%
\sum_{i=1}^{M}\left| w_{i}\right\rangle \text{,} \\
\left| \underset{\symbol{126}}{r}\right\rangle &=&\frac{1}{\sqrt{N-M}}%
\sum_{i=M+1}^{N}\left| w_{i}\right\rangle \text{ .}
\end{eqnarray*}%
Clearly, now both $\left| \underset{\symbol{126}}{w}\right\rangle $ and $%
\left| \underset{\symbol{126}}{r}\right\rangle $ are unit and orthogonal to
each other.

Next, let the Grover kernel $G=-G_{2}G_{1}$ operate on $\left|
s\right\rangle $. Four phase parameters $\alpha $, $\beta $, $\gamma $, and $%
\delta $ are given in the two Grover operators $G_{1}$ and $G_{2}$
respectively, such that in the space spanned by $\left| \underset{\symbol{126%
}}{w}\right\rangle $ and $\left| \underset{\symbol{126}}{r}\right\rangle $,
we have

\begin{eqnarray*}
G_{1} &=&\alpha \left| \underset{\symbol{126}}{w}\right\rangle \left\langle 
\underset{\symbol{126}}{w}\right| +\beta \left| \underset{\symbol{126}}{r}%
\right\rangle \left\langle \underset{\symbol{126}}{r}\right| \\
&=&\left[ 
\begin{array}{cc}
\alpha & 0 \\ 
0 & \beta%
\end{array}%
\right] \text{,}
\end{eqnarray*}

\begin{eqnarray*}
G_{2} &=&\gamma \left| s\right\rangle \left\langle s\right| +\delta
(I-\left| s\right\rangle \left\langle s\right| ) \\
&=&\left[ 
\begin{array}{cc}
\delta +(\gamma -\delta )\frac{M}{N} & (\gamma -\delta )\frac{\sqrt{M(N-M)}}{%
N} \\ 
(\gamma -\delta )\frac{\sqrt{M(N-M)}}{N} & \gamma -(\gamma -\delta )\frac{M}{%
N}%
\end{array}%
\right] \text{.}
\end{eqnarray*}%
Consequently, the Grover kernel then is given by

\begin{equation}
G=-\left[ 
\begin{array}{cc}
\alpha (\delta +(\gamma -\delta )\frac{M}{N}) & \beta (\gamma -\delta )\frac{%
\sqrt{M(N-M)}}{N} \\ 
\alpha (\gamma -\delta )\frac{\sqrt{M(N-M)}}{N} & \beta (\gamma -(\gamma
-\delta )\frac{M}{N})%
\end{array}%
\right] \text{.}
\end{equation}%
Note that alternatively the parameters are written

\begin{equation}
\alpha =e^{i\theta _{1}}\text{, }\beta =e^{i\theta _{2}}\text{, }\gamma
=e^{i\phi _{1}}\text{ and }\delta =e^{i\phi _{2}}  \notag
\end{equation}%
where the phases $\theta _{1}$, $\theta _{2}$, $\phi _{1}$, and $\phi _{2}$
are for the marked state, the state orthogonal to the marked state, the
initial state, and the state orthogonal to the initial state, respectively.

Finally, repeat $m$ times the operation, and wish the probability of finding
the marked element $\left| \underset{\symbol{126}}{w}\right\rangle $ be
greater than some particular value. We then require, say,

\begin{equation}
p=\left| \left\langle \underset{\symbol{126}}{w}\right| G^{m}\left|
s\right\rangle \right| ^{2}>\frac{1}{2}\text{.}
\end{equation}%
The amplitude $\left| \left\langle \underset{\symbol{126}}{w}\right|
G^{m}\left| s\right\rangle \right| $ is deduced

\begin{equation}
\left\langle \underset{\symbol{126}}{w}\right| G^{m}\left| s\right\rangle
=\xi _{2}^{m}(\sqrt{\frac{M}{N}}+(\xi _{1}/\xi _{2}-1)\langle \underset{%
\symbol{126}}{w}|g_{1}\rangle \left\langle g_{1}|s\right\rangle )\text{,}
\end{equation}%
where $\xi _{1,2}$ and $\left| g_{1,2}\right\rangle $ denote the eigenvalues
and the corresponding eigenvectors of $G$, respectively. The detail
expression of the eigenvalues and eigenvectors are given by

\begin{eqnarray}
\xi _{1,2} &=&e^{i\lambda _{1,2}}=\frac{1}{2}\text{Tr}G\pm \frac{1}{2}\sqrt{(%
\text{Tr}G)^{2}-4\text{Det}G}\text{,} \\
\left| g_{1,2}\right\rangle &\varpropto &\left[ 
\begin{array}{c}
\frac{k\pm N\sqrt{(\text{Tr}G)^{2}-4\text{Det}G}}{2\alpha (\gamma -\delta )%
\sqrt{M(N-M)}} \\ 
1%
\end{array}%
\right] \text{,}
\end{eqnarray}%
where

\begin{eqnarray}
\text{Tr}G &=&-\frac{M(\alpha -\beta )(\gamma -\delta )+N(\gamma \beta
-\alpha \delta )}{N}\text{,} \\
\text{Det}G &=&\alpha \beta \gamma \delta \text{,}  \notag \\
k &\equiv &N(\gamma \beta -\alpha \delta )-M(\alpha +\beta )(\gamma -\delta )%
\text{.}  \notag
\end{eqnarray}%
The eigenvector $\left| g_{1}\right\rangle $ should be discussed in detail
because, as shown in (5), it determines whether the requirement (4) can be
satisfied. As $M<<N$ and $N>>1$, the eigenvector $\left| g_{1}\right\rangle $
can be asymptotically expressed by

\begin{equation}
\left| g_{1}\right\rangle \varpropto \left[ 
\begin{array}{c}
\frac{\alpha \delta -\beta \gamma }{\alpha (\gamma -\delta )}\sqrt{\frac{N}{M%
}} \\ 
1%
\end{array}%
\right] \sim \left| \underset{\symbol{126}}{w}\right\rangle
\end{equation}%
and then $\langle \underset{\symbol{126}}{w}|g_{1}\rangle \left\langle
g_{1}|s\right\rangle =O(\frac{1}{\sqrt{N}})$, meaning that the requirement
(4) will never be satisfied. Figure 1 shows an example for this case. To
avoid (8), we in turn must have the matching condition

\begin{equation}
\frac{\alpha }{\beta }=\frac{\gamma }{\delta }\text{ or }\theta _{1}-\theta
_{2}=\phi _{1}-\phi _{2}
\end{equation}%
such that the eigenvector then becomes, in the normalized form

\begin{equation*}
\left| g_{1}\right\rangle =\left[ 
\begin{array}{c}
\sqrt{\frac{(\alpha -\beta )\delta }{(\gamma -\delta )\alpha }} \\ 
1%
\end{array}%
\right] /\sqrt{2}\text{.}
\end{equation*}%
We then have, under the matching condition (9),

\begin{equation*}
\langle \underset{\symbol{126}}{w}|g_{1}\rangle \left\langle
g_{1}|s\right\rangle =(\sqrt{\frac{M}{N}}+\sqrt{\frac{N-M}{N}}%
)/2\thickapprox \frac{1}{2}\text{.}
\end{equation*}%
Now, since

\begin{equation*}
\left| \xi _{1}/\xi _{2}-1\right| =\left| e^{im\Delta \lambda }-1\right|
=2\left| \sin \frac{m\Delta \lambda }{2}\right| \text{,}
\end{equation*}%
where $\Delta \lambda =\lambda _{1}-\lambda _{2}$, the probability then is

\begin{eqnarray}
p &=&\left| \left\langle \underset{\symbol{126}}{w}\right| G^{m}\left|
s\right\rangle \right| ^{2}=\left| \sqrt{\frac{M}{N}}+(e^{im\Delta \lambda
}-1)\underset{\symbol{126}}{\langle w}|g_{1}\rangle \left\langle
g_{1}|s\right\rangle \right| ^{2}  \notag \\
&\approx &\left| \sin \frac{m\Delta \lambda }{2}\right| ^{2}\text{.}
\end{eqnarray}%
Figure 2 shows the $p$ vs. $m$ diagram under the condition (9). It shows
that $p$ can approach unity as $\left| m\Delta \lambda \right| =(2j-1)\pi $, 
$j=1,2,...$. Taking first $\pi $ , we have%
\begin{eqnarray}
m &=&\left| \frac{\pi }{\Delta \lambda }\right|  \notag \\
&=&\frac{\pi }{2}\sqrt{\frac{M}{N}}\left| \frac{\alpha \delta }{(\alpha
-\beta )(\gamma -\delta )}\right| ^{\frac{1}{2}}  \notag \\
&=&\frac{\pi }{2}\sqrt{\frac{M}{N}}\left[ \frac{1}{2\cos (\phi _{1}-\phi
_{2})}\right] ^{\frac{1}{2}}
\end{eqnarray}%
Eventually, the searching time step will be minimum when

\begin{equation*}
\theta _{1}-\theta _{2}=\pi \text{,}
\end{equation*}%
and the corresponding time step is

\begin{equation*}
\text{min}(m)=\frac{\pi }{4}\sqrt{\frac{M}{N}}\text{.}
\end{equation*}%
In figure 3, the $p$ vs. $m$ diagram is shown under an optimal choice of the
parameters.

To summarize, as the four parameters $\alpha $, $\beta $, $\gamma $, and $%
\delta $ are taken into consideration in a generalized Grover's kernel $G$,
the phase matching condition for these parameters has been deduced. It is
found that the marked elements will never be searched unless $\alpha \delta
-\beta \gamma =0$. That is, the eigenvector $\left| g_{1}\right\rangle $ of
the Grover kernel $G$ should not asymptotically coincide the marked state.
The optimal option for the relation between these parameters, however, is

\begin{equation}
\frac{\alpha }{\beta }=\frac{\gamma }{\delta }=1\text{ or }\theta
_{1}-\theta _{2}=\phi _{1}-\phi _{2}=\pi 
\end{equation}%
The choice of parameters taken in the original Grover operator is only the
simplest one, in which $\alpha =\gamma =-1$ and $\beta =\delta =1$ were
used. Long et al. \cite{long1} have treated the phase rotations for the
initial and the marked states alone, and the phase matching condition is
exactly the case of $\alpha =\gamma $ and $\beta =\delta =1$. Galindo and
Mart\'{\i}n-Delgado\cite{galindo}, on the contrary, have discussed the phase
rotations for the states orthogonal to the initial and the marked states,
and have addressed the condition $\alpha =\gamma =-1$ and $\beta =\delta $.

\bigskip

{\LARGE Figure Captions}

Fig.1 Probability $p$ as a function of the time step $m$ for $N=1000$, $M=10$%
, $\alpha =\gamma =e^{i\pi }$, $\beta =e^{i\frac{\pi }{2}}$, $\delta
=ie^{i3} $.

Fig.2 Probability $p$ as a function of the time step $m$ for $N=1000$, $M=10$%
, $\alpha =e^{i1.7\pi }$, $\gamma =e^{i\pi }$, $\beta =e^{i1.6\pi }$, $%
\delta =e^{i0.9\pi }$.

Fig.3 Probability $p$ as a function of the time step $m$ for $N=1000$, $M=10$%
, $\alpha =e^{i1.7\pi }$, $\gamma =e^{i1.9\pi }$, $\beta =e^{i0.7\pi }$, $%
\delta =e^{i0.9\pi }$.

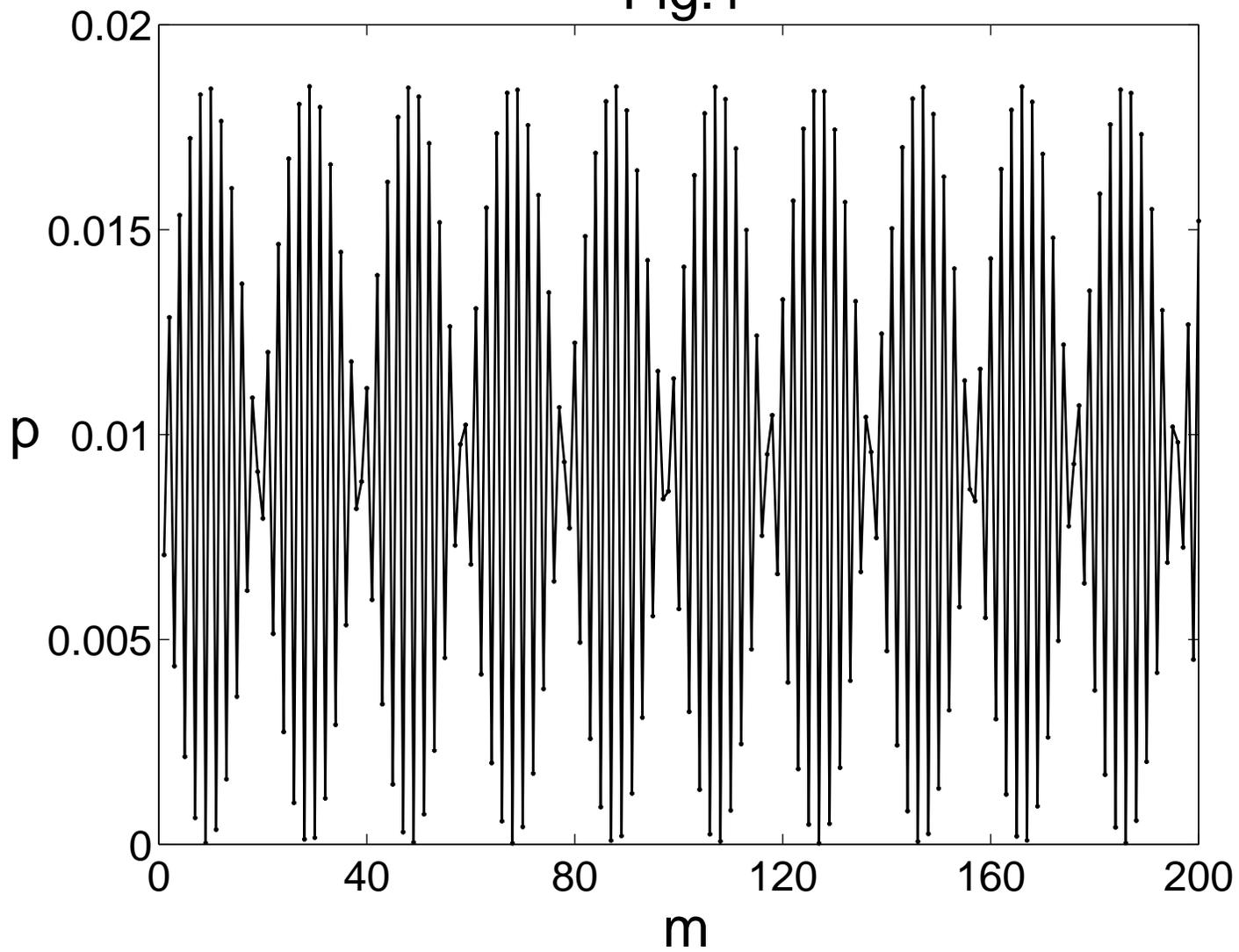
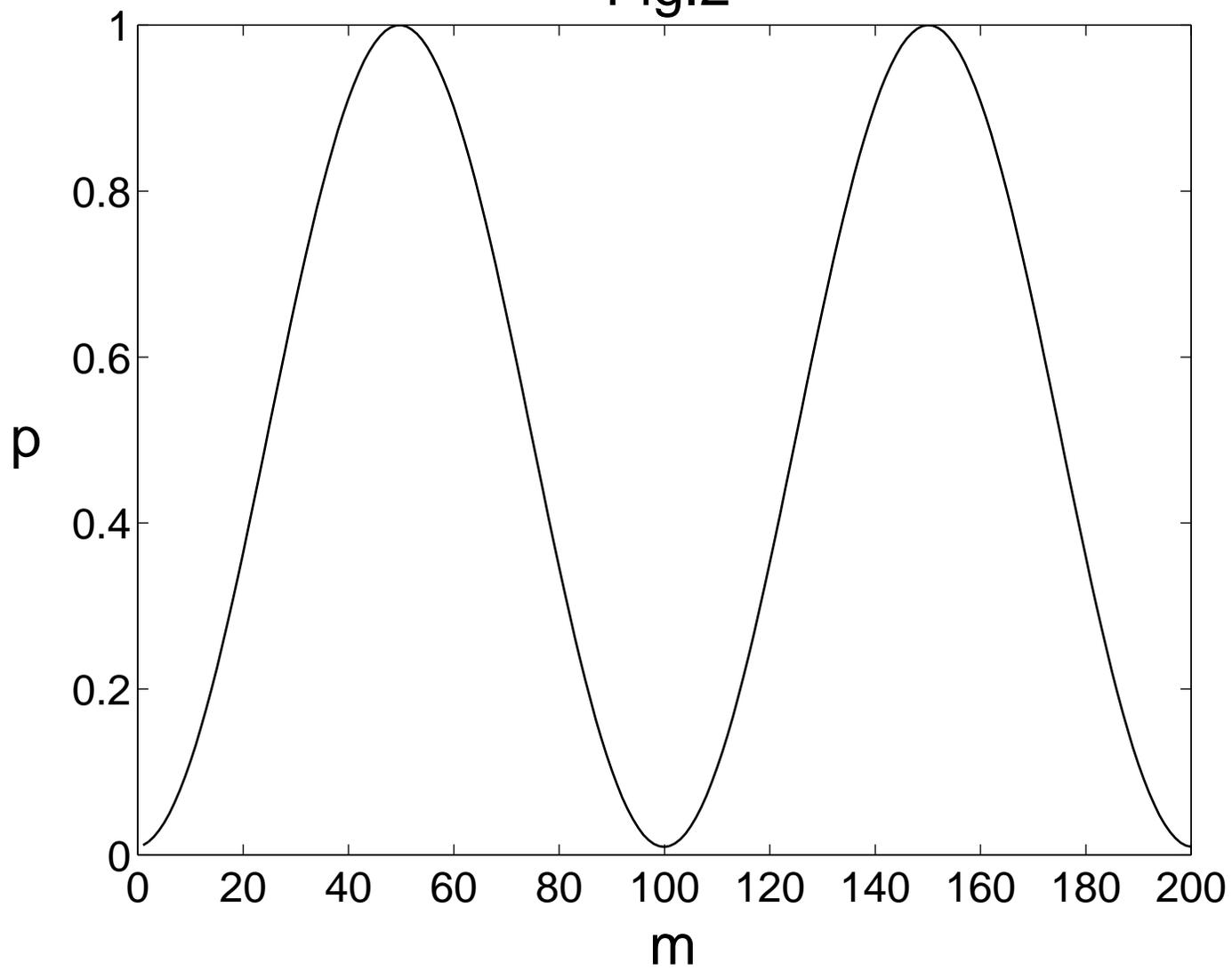
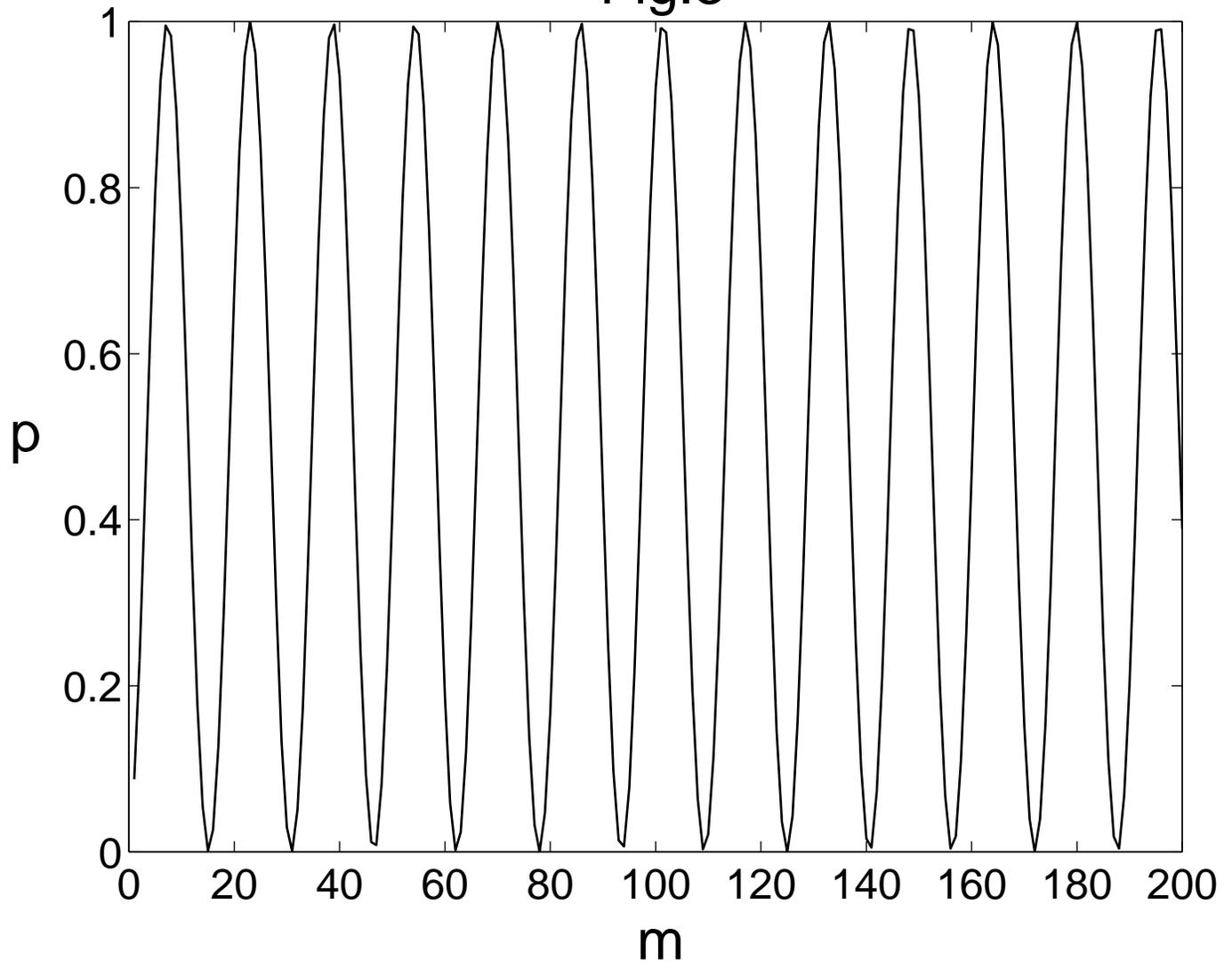
\end{document}